\documentstyle[11pt,psfig]{article}
\begin{document}
\date{}
\title{From Quasars to Extraordinary N-body Problems}
\author{D. Lynden-Bell}
\maketitle
\centerline{Clare College, Cambridge \& Institute of Astronomy.}
\centerline{Visiting Professorial Fellow, The Queen's University, Belfast.}

\begin{abstract}
We outline reasoning that led to the current theory of quasars and
look at George Contopoulos's place in the long history of the N-body
problem.  Following Newton we find new exactly soluble N-body problems
with multibody forces and give a strange eternally pulsating system
that in its other degrees of freedom reaches statistical equilibrium.
\end{abstract}

\section*{Introduction}

This meeting is held to honour George Contopoulos for his great
contributions to dynamical systems theory and the N-body problem.  I
shall pay my tribute to him in three parts,

\begin{itemize}

\item[I] {Placing his contributions in the proud history of those who have
made major contributions to the N-body problem.}

\item[II] {Since nothing but the best is good enough to honour George
I present to him a copy of my best paper (Lynden-Bell, 1969) and
include here a r\'esum\'e of its arguments that led to the current
theory of quasars.  There are questions my paper raised 29 years ago
which are still unexplored.}

\item[III] {With Prof. Ruth Lynden-Bell (my wife) I present our new
extraordinary N-body problems which we solve for all initial
conditions.  These problems can also be solved in quantum mechanics
when the hyper-keplerian potential energy is $$V = -N \widetilde Z e^2/r \
, \eqno (1)$$
where
$$r^2 = \sum ^N _ {i=1} {m_i \over M} \left ( {\bf x}_i - {\overline
{\bf x}} \right) ^2 \ , \eqno (2)$$

$$M = \sum m_i \eqno (3)$$
i.e., $r$ is the mass-weighted-root-mean-square radius of the N-body
system.  I quote here the energy and degeneracy of the $n$th quantum
state but we shall publish derivations elsewhere.
$$E_n = - {2M \hbar^{-2} \left( N {\widetilde Z} e^2 \right) ^2 \over \left[ 2n + 3 (N-2) \right] ^2} \eqno (4)$$
the degeneracy of this $N$ particle state is
$$g(n, N) = {\left [ n+3 (N-2) \right] ! \over (n-1)! (3N-4)!} \left
[ 2n + 3 (N-2) \right] \ ,  \eqno (5)$$
for $N=2$, $g$ reduces to $n^2$ and the energy reduces that of the
hydrogen atom for which $M = m_p + m_e$.  To see this, $\widetilde Z$ is
replaced by ${\textstyle {1 \over 2}} (m_p m_e)^{1/2}/M$ when $r$ is
replaced by the separation of the electron from the proton.  We have
written the coefficient of the potential energy in the clumsy form $N
\widetilde Z e^2$ so that the analogy to hydrogenic atoms can be readily
seen by physicists.
}
\end{itemize}

\section{Contributions to the N-body problem (excluding
Agatha Christie's)}

The N-body problem probably started with Newton although Hooke would
undoubtedly dispute it since he seems to have conceived the idea
independently but had not the mathematical ability to work out its
consequences.  As Chandrasekhar (1996) has shown, Newton's Principia
(Newton 1687, Cajori 1934) has much to teach us even today (see
Lynden-Bell \& Nouri-Zonoz, 1998).  Recent studies of the Portsmouth
papers have shown that Newton developed most of the perturbation
theory that was hitherto attributed to the mathematical astronomers of
the 18th and 19th centuries.  Newton's method was to store up the
momentum generated by perturbations and then deliver it as an impulse
that changed the motion from one ellipse to another.  This of course
gives him the equations for the variations of the orbital elements
which are the meat of perturbation theory.  My brief r\'esum\'e of the
N-body problem's history is:

\medskip

\begin{tabular}{ll}
Newton 1687 & Orbit Theory and the general solution of \\
&
the first extraordinary N-body problem \\
&\\
Laplace 1795 & Perturbation Theory for near circular \\
& orbits \\
& \\
Poincar\'e 1892--99 & Topological Methods\\
&\\
Whittaker 1913 (1959) & Adelphic Integrals as series \\
&\\
Contopoulos 1956--  & Third Integrals and Chaos \\
&\\
Kolmagorov-Arnold-Moser & Invariant Tori \& Arnold diffusion \\ 

\end{tabular}

\bigskip

To these theoretical studies we must add the numerical computation of
the N-body problem and here Aarseth's name stands out as a persistent
pioneer exploring this problem (Aarseth, 1974) although many others
have contributed, especially Heggie (1975) through his work on triple
interactions.  Both in globular cluster theory and in dynamical
systems Henon's work stands out for its beauty (Henon, 1961, 1969,
1974) while Antonov (1962) was responsible for a fundamental advance
in the understanding of gravitational thermodynamics later popularised
and extended to negative specific heats and gravitational phase
transitions by the author (Lynden-Bell \& Wood, 1968; Lynden-Bell \&
Lynden-Bell, 1977)and by Thirring (1972).  Betteweiser \& Sugimoto
(1984) were responsible for giving the gravothermal instability a
delightful new twist in their discovery of the inverse gravothermal
catastrophe that leads to giant thermal oscillations.  But let me
return to what George Contopoulos taught me at our many contacts since
1961.

To set the scene I had written my thesis in 1960 which contained a new
derivation of what potentials had local first integrals of the motion
besides the energy and the angular momentum about the axis.  The main
part of the work was the derivation of these different classes of
potential while other parts of the thesis contained the time dependent
evolution of accretion disks\footnote{they got that name only later}
and a first attempt to apply Jeans's (1928) gravitational instability to make
a theory of the spiral structure of galaxies.  The beliefs of those
times are well illustrated by the first edition of Landau \&
Lifshitz's book on classical mechanics; either a dynamical system was
separable and integrable or it was ergodic (by which was meant that
almost all orbits visited all volumes of the phase space accessible
under the energy constraint).  Having classified the special forms of
potential that had local integrals I expected that most other
potentials would show ergodic behaviour.  From the inequality of the
$z$ and $R$ dispersions of the stars in the Galaxy it was clear that
there must be another integral other than $E $ and $h$ for the Milky
Way so I had begun trying to fit Eddington (1915) (now called Stakle)
potentials to galaxies.  It was quite shattering when at the 1961 IAU
general assembly in Berkeley, George Contopoulos (1960) showed that orbits in
most smooth potentials behaved as though there were third integrals.
Suddenly the special interest of the special potentials fell away ---
they were not the only systems with 3rd integrals, merely those for
which we knew the exact analytical form of those integrals.  They
seemed now to be mathematical curiosities rather than systems
fundamental to the dynamics of real galaxies.  

Three years later George organised a very instructive IAU symposium
(No.25) at Thessaloniki on the Theory of Orbits in the Solar System
and Stellar Systems.  Here he brought into contact the celestial
mechanics fraternity, with their long history of analytically
calculating orbits in the solar system by perturbation theory, with us
new boys who were attempting to understand the statistics of the
orbits in the more complicated potentials of galaxies; George
Contopoulos (1965, 1966) here taught us that many of the problems were
common to both fields and showed how fertile it was to bring different
communities who knew different things to the same conference -- his
wide interests have made him especially good at that throughout his
life and this 1998 conference is no exception.

For brevity, I shall skip contacts at Besan\c{c}on on the N-body problem
where George presented Poisson Bracket  series for third integrals and we
were introduced to Lie series.

In 1973 at Saas F\'ee George gave lectures in which he introduced me
to the wonders of modern dynamical theory -- topological methods
incomplete chaos and the KAM theorem.  It opened my eyes to so much
that was new to me that I retreated back to more directly astronomical
topics preferring the contact with astronomy to the unchartered seas
revealed by this new alliance between computers and topology.

Two years ago at Salsjobaden in a conference on the Dynamics of Barred
Spirals, George again broke open a new field (Contopoulos, 1997).  His
invariant dynamical spectra (described also in his contribution here)
taught us how to measure and classify chaos, even complete chaos!

I have picked out a tiny fraction of George Contopoulos's work (1975) and
mentioned things I learned from our direct contacts.  He will no doubt
deduce that I am not a very attentive pupil but it would be mean not
to  mention a lovely paper on the light distributions of elliptical
galaxies (Contopoulos, 1956) because it is a beautiful work to which I
constantly have to refer my astronomical colleagues!

The essence of this paper can be deduced by the following argument.
Consider a spherical galaxy with any radial light profile.  Now
flatten its density distribution by linear contraction along any
axis.  This contraction can be resolved into one along and one
perpendicular to the line of sight.  The one along makes no difference
while the one perpendicular flattens the circular distribution of
observed light into one stratified on similar ellipses.  If a further
contraction is made along another axis we can apply the same argument
again since ellipses contracted along any direction remain ellipses.
So we arrive at George's beautiful theorem that if the density
distribution of an elliptical galaxy is stratified on similar
concentric ellipsoids then the light seen will be stratified on
similar concentric ellipses whatever the orientation of the galaxy to
the line of sight.

\section{Background to the Accretion Disc Theory of Quasars}

My own best work is ``Galactic Nuclei as Collapsed Old Quasars''
written in 1969.  Then the discovery of quasars by Schmidt using
Hazard's accurate position for one of Ryle's radio sources was still
recent and quasars themselves were enigmatic objects more especially
so because even the brightest 3C273 and 3C48 too did not seem to be
associated with clusters of galaxies.  

No-one then knew that Michell in his wonderfully percipient paper of
1784 had predicted both giant black holes and how they would be
discovered!  Even the name black hole only came into general use in
1970!  In my 1969 paper I refer to ``Schwarzschild throats''.
Laplace's translation of Michell's work into French (without
attribution!) was not common reading among astronomers either.

Among the modern works on quasars as accretion discs, priority goes to
Salpeter's fine 1964 letter to the Astrophysical Journal.  Turning
against the then common view that quasars were not associated with
clusters of galaxies he worked out the consequences of a large black
hole moving through a galaxy and accreting according to the Hoyle \&
Lyttleton formula.  He derived the power emitted per unit accretion
rate by considering the binding energy of the last stable circular
orbit and deduced a number of consequences of such black holes
accreting as they wandered through the interstellar gas of a galactic
disc.  Five years elapsed before I wrote my paper.  Originally unaware
of Salpeter's (1964) note I luckily learned of it before the proofs came and
so was able to add a sentence and a reference to his work.  My aim was
to show that the very small nuclei already known in the centres of
galaxies were likely to be stars gathered around the giant-black-hole
remnants of quasars.  At the time, 1969, we already knew that the
Optical Violently Variable or OVV quasars could change by a magnitude
from one night to the next.  Geoffrey Burbridge (1958) had been insistent
that the giant radio sources needed $10^{61}$ ergs in fast electrons
and magnetic field, while Ryle (1968) had emphasised that quasars would
not be distinguished from such sources by radio measurements.

Now $10^{61}$ ergs weigh ${\textstyle {1 \over 2}} 10^7$M$_\odot$.  If
one entertained the idea that these ergs came from nuclear energy then
the 1\% mass conversion efficiency of nuclear burning means that
$10^9$M$_\odot$ are needed.  However putting $10^9$M$_\odot$ within
the light-variation-time length-scale of 10 light hours gives a
gravitational binding energy of $10^{62}$ ergs -- on such a hypothesis
$10^{62}$ ergs of gravitational energy would have been lost, all in
order to burn $10^9$M$_\odot$ of hydrogen into Helium and thereby get
the mere $10^{61}$ ergs needed.  This shows that in assuming nuclear
power we nevertheless conclude that most of the energy comes from
gravity.  So the nuclear idea is not sensible and we should assume a
preponderant gravity power and a somewhat smaller mass $\sim
10^8$M$_\odot$.  If conversion of mass into radiation is not 100\%
efficient quasars must leave behind massive remnants of $>10^7 -
10^8$M$_\odot$ and because they have radiated their binding energy
they have insufficient energy to re-expand.  Since the masses are far
beyond the Chandrasekhar limit there are no other final resting places
other than giant black holes.  Turning to the numbers of quasars
derived by Sandage and estimating possible lifetimes, I deduced
$${{\rm Number \ of \ clusters} \atop {\rm of \ galaxies}} \quad <
\quad {{\rm Number \ of \ dead} \atop {\rm quasars}}\quad < \quad
{{\rm Number \ of} \atop {\rm galaxies}}$$ 

\begin{figure}[th]
\psfig{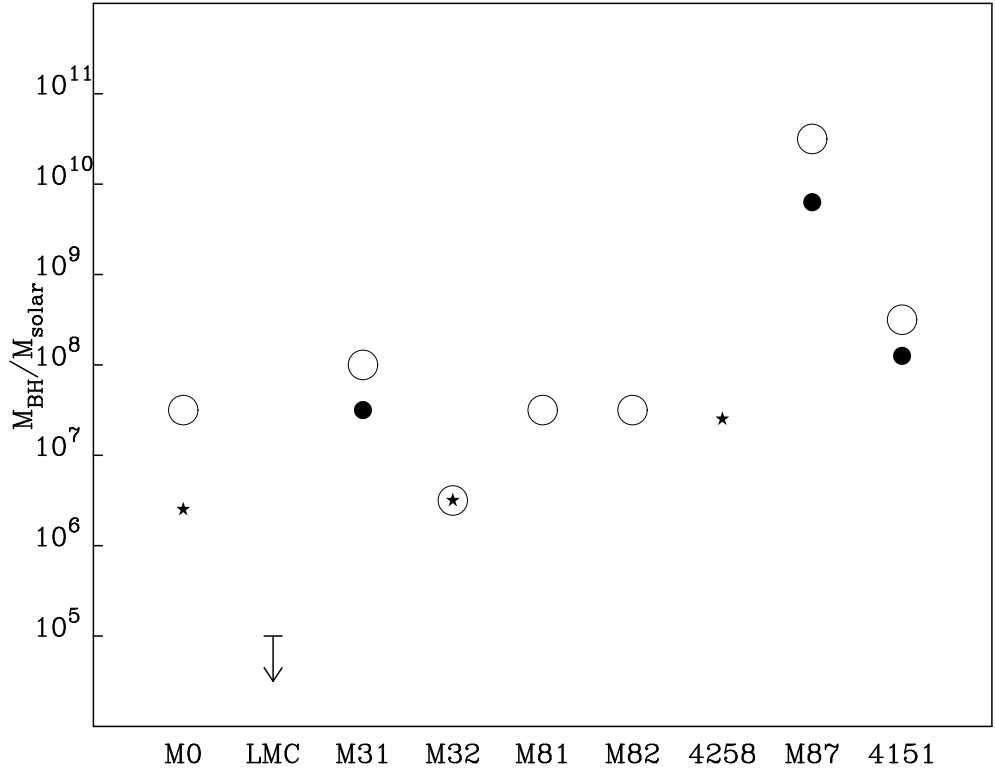}
\caption{Comparison of my 1969 estimates of black hole masses $\bigcirc$
 with modern determinations $\star$ and less precise modern estimates
$\bullet$.  Evidently the early estimates were somewhat over
enthusiastic!}
\end{figure}
\noindent
Thus the nearest dead
quasar must be nearer than M87 and there may be as many dead quasars
as there are massive galaxies.  How could we hide dead quasars of
$10^8$M$_\odot$ when they still gravitate?  They would naturally be
centres of attraction for stars so it is natural to find such a body
at the centre of an exceptionally dense region.  Galactic Nuclei then
became the obvious candidates so I looked at the Galaxy, the Magellanic
Clouds, M31, M32, M81, M82, NGC4151, M87, etc. and estimated possible black
hole masses from the 1969 data on their nuclei, many of which were due
to pioneering work by Merle Walker, see Figure 1.  I also drew on the
accretion discs of my thesis and, finding the gaseous viscosity too
low, I estimated a magnetic viscosity.  This was based on the shearing
of the disc causing magnetic reconnection and continual flaring above
the disc.  Indeed I found that the protons got most of the energy as
they more readily achieved ``runaway''.  Particle energies up to
$10^{13}$eV were readily generated and hard emission would result when
this hit the disc material.  While the energy was primarily dissipated
into such fast cosmic rays they would collide with the disc and heat
it to temperatures $T \propto r^{-3/4}$ for $r \gg 2GM_0/c^2$.  Adding
together such black body rings of emission, I got the disc spectrum
$S_\nu \sim \nu^{1/3}\, \exp - (h \nu/kT_{\max})$ where $T_{\max}$ the
maximum temperature in Kelvin was $6.6 \times 10^4 F_{-3}^{1/4}
M_7^{-1/3}$; here $F_{-3}$ is the mass flux in units of
$10^{-3}$M$_\odot$/yr with $M_7$ the black hole's mass in units of
$10^7$M$_\odot$.  {\em I did not estimate how much hard emission would
come from the initial collisions of the cosmic rays with the disk but
a $10^{13}$ {\rm eV} cosmic ray is certainly capable of emitting hard
$\gamma$ rays at its first few collisions.}  Even today, 29 years
later, I think this model deserves more attention as a serious rival
to the currently popular advection models.

  The following year Jim Bardeen wrote a particularly
fine paper which showed how accretion would spin up a Schwarzschild
hole and after a finite mass was accreted leave it growing as a
near-limiting Kerr hole of significantly greater efficiency.  I gave a
paper on these models at the 1970 Vatican Symposium on the nuclei of
galaxies and in 1971 re-estimated the luminosity functions of quasars
and mini quasars by developing the C$^-$ method.  That year Ekers was
stimulated to look with higher radio resolution at the Galactic Centre
(Ekers \& Lynden-Bell, 1971) and I reviewed the then known data on it
with Rees (Lynden-Bell \& Rees, 1971).  A year or two later attention
turned to lower mass black holes with the discovery of many X-ray
binaries by the UHURU satellite.  Papers by Pringle \& Rees (1972) and
Shakura \& Sunyaev (1973, 1976) applied such ideas on a smaller scale
and with Pringle I applied them (1974) to star formation both with and
without magnetic fields.  In 1978 I introduced the thick Kerr and
Schwarzschild vortices in the hope of getting a more natural
collimation mechanism than that of Blandford \& Rees (1974) but the
very narrow jets are still inadequately understood.

\section{General Exact Solution to an Extraordinary N-body Problem}
I now return to the N-body problem and the little known fact that in
Principia Newton (1687) solved an N-body problem in which every body
attracts every other one and he solved it for all initial conditions!

His was the first of the class of extraordinary N-body problems which
Ruth Lynden-Bell and I have been studying.  Newton took the force
between two bodies $i$ and $j$ to be $F = k m_i m_j ({\bf x}_j -
{\bf x}_i)$.  To get the total force in particle $i$ he summed over $j$ and
since the $j=i$ term is zero we may sum over all $j$ to obtain
$${\bf F}_i = \sum _j {\bf F}_{ij} = k m_i M({\overline {\bf x}} -
{\bf x}_i) \eqno (6)$$
where ${\overline {\bf x}}$ is the position vector of the centre of
mass which of course moves uniformly in a straight line and $M$ is the
total mass of the system.  Thus with this linear mass-weighted law,
that Newton would never have ascribed to Hooke, the total force on the
$i^{{\rm th}}$ body is directed to the centre of mass and proportional
to the distance from it.  Therefore Newton found that each body
describes a centred ellipse about the centre of mass which itself
moves uniformly.  This completes Newton's solution.  In his case the
potential energy is
$$V = - {\textstyle {1 \over 2}} K \sum_{\ \ i \ <} \sum_{\! \! \! \!
j} m_i m_j ({\bf x}_i - {\bf x}_j)^2 = -{\textstyle {1 \over 2}} k M
\sum_i m_i ({\bf x}_i - {\overline {\bf x}} ) ^2 = -{\textstyle {1
\over 2}}k M^2 r^2 \ . \eqno (7)$$

Generalising some work on statistical mechanics by Ruth Lynden-Bell we
were led to consider the dynamics of N-body systems with the more
general potential energy $V= V(r)$ where $r$ is given above.
(cf. equation (2)).  We define a mass weighted radius ${\bf r}$ in
3$N$ dimensions by
$${\bf r} = \left( \sqrt{{m_1 \over M}} \left( {\bf x}_1 - {\overline
{\bf x}} \right) \ , \ \sqrt{{m_2 \over M}} \left( {\bf x}_2 -
{\overline {\bf x}} \right) \ , \ \sqrt{{m_N \over M}} \left( {\bf x}_N
- {\overline {\bf x}} \right) \right) \ , \eqno (8)$$
so the first 3 of the $N$ coordinates tell us where particle 1 is,
the next 3 where particle 2 is, etc.  Notice that $|{\bf r}|$ is the
$r$ we defined previously.  Equations of motion of the particles in
centre of mass coordinates then lead directly to the equation
$$M{\ddot {\bf r}} =-V'(r) {\hat {\bf r}} \eqno (9)$$
where ${\hat {\bf r}} = {\bf r}/r$ is the unit radial vector in 3$N$
space.

One readily sees that
$$r_\alpha {\ddot r} _\beta - r_\beta {\ddot r}_\alpha = 0 \ , \eqno
(10)$$
so
$$r_\alpha {\dot r}_\beta - r_\beta {\dot r}_\alpha = L_{\alpha \beta}
= -L_{\beta \alpha} = \ {\rm const} \ . \eqno (11)
$$
Furthermore
$$L^2 = {\textstyle {1 \over2}}L_{\alpha \beta} L_{\alpha \beta} = {\textstyle {1 \over 2}}\left( r_\alpha {\dot
r}_\beta - r_\beta \dot r_\alpha \right)  \left(r_\alpha \dot r_\beta
- r_\beta \dot r_\alpha \right) =  \left[ r^2 ({\dot {\bf r}})^2 -
({\bf r} \cdot {\dot {\bf r}})^2 \right] \eqno (12)$$
where $L^2$ is the constant defined by the first equality.

The energy in centre of mass coordinates is given therefore by 
$${\textstyle {1 \over 2}} M {\dot {\bf r}}^2 + V(r) = E = {\textstyle
{1 \over 2}} M({\dot r}^2 + L^2 r^{-3}) + V(r) \ . \eqno (13)$$
This determines $r(t)$ as a periodic function if $E<0$ so there is no
violent relaxation in these systems and they vibrate eternally.  

Differentiating (13) we find
$$M (\ddot r - L^2 r^{-3}) = -V' (r) \ . \eqno (14)$$
This is the same equation of motion as that for the central distance
to an object in planar motion which angular momentum $L$ about a
centre of force with potential $V(r)$.  It is natural to imagine such
a planar orbit and to invent an angle $\phi$ such that $\phi = 0$ at
some pericentre and 
$$r^2 {\dot \phi} = L \ , \eqno (15)$$
we may then imagine an orbit in two dimensional polar coordinates $r,
\phi$ and following Newton we shall cling to the geometry by
eliminating the time in favour of $\phi$.  Now
$${\ddot {\bf r}} = {d^2 \over dt^2} (r {\hat {\bf r}}) = {d \over dt}
\left( \dot r {\hat {\bf r}} + {L \over r} {d {\hat {\bf r}} \over d \phi}
\right) = \ddot r {\hat {\bf r}} + {L^2 \over r^3} \, {d^2 {\hat {\bf r}}
\over d \phi^2} \eqno (16)$$
where two terms in ${\dot r} L r^{-2} d{\hat {\bf r}}/d\phi$ cancel at the
last step.  Inserting this result into our equation of motion (9) and
using (14), we deduce the wonderfully simple equation
$$d^2 {\hat {\bf r}}/d \phi^2 + {\hat {\bf r}} = 0 \ , \eqno (17)$$
whose solution is 
$${\hat {\bf r}} = {\bf A} \cos \phi + {\bf B} \sin \phi \eqno (18)$$
where $\bf A$ and $\bf B$ are constant 3$N$-vectors which obey $A^2 =
B^2 = 1$ and ${\bf A} \cdot {\bf B} = 0$ in order that ${\hat {\bf
r}}$ should be a unit vector for all $\phi$.  Three further
constraints on ${\bf A}$ and ${\bf B}$ follow from the fixed centre of
mass.  They are detailed in our paper but need not concern us here.

We now have the general solution, the centre of mass moves uniformly
in a line and the particles pursue orbits about it of the form
$${\bf r} = r (\phi) ({\bf A} \cos \phi + {\bf B} \sin \phi) \ , \eqno
(19)$$
where $r(\phi)$ is the form of the two dimensional orbit governed by
equations (13) and (15).  These can be integrated explicitly for the
Isochrone potential $V \propto k/(b+s)$,  $s^2 = r^2 + b^2$ and for
the Kepler and harmonic oscillator potentials.  For the Kepler case
$r(\phi) = \ell / (1 + e \cos \phi)$ so the solution is of the
pleasing form
$${\bf r} = \ell (1 + e \cos \phi)^{-1} ({\bf A} \cos \phi + {\bf B}
\sin \phi) \ .$$
If we concentrate on the particle $i$, we find its orbit lies in the plane
perpendicular to ${\bf A}_i \times {\bf B}_i$ where $i$ denotes the
three components corresponding to particle $i$.  Taking $x, y$
coordinates in that plane and eliminating $\phi$ we find that the
orbit is quadratic.  If $e$ were zero it would be a central ellipse,
while if $|{\bf A}_i|$ and $|{\bf B}_i|$ are equal and orthogonal it
gives a Keplerian eccentric ellipse.  In the general bound case the
ellipse has neither its centre nor its focus at the centre of mass $r
= 0$.   These systems obey the equilibrium Virial theorem in the form
$2 {\cal T} - r V' = 0$, so for the hyper-keplerian case $V \propto
r^{-1}$ it takes the more familiar form $2 {\cal T} + V = 0$.

One may work out the microcanonical statistical mechanics and find
that \break $E = -{\textstyle{3 \over 2}}(N-2) kT$ so that the heat capacity $C = - {\textstyle{3
\over 2}}(N-2)k$ which is clearly negative as for other gravitating
systems (Lynden-Bell \& Wood, 1968) and black holes.  If $V$ takes the form
$$ V = \left\{  
\begin{array}{ll}
\infty & \qquad r<b \\
-kM^2/r & \qquad b < r < R \\
\infty & \qquad r>R 
\end{array} 
\right . $$ corresponding to a gravitating system which cannot get too
small or too big then a Canonical ensemble is possible and the
negative specific heat region of the microcanonical ensemble is
replaced by a giant first order phase transition as in our earlier
model (Lynden-Bell \& Lynden-Bell, 1977).

\section*{3b  Generalisation}

We may extend these extraordinary N-body problems by taking $V$ to be
of the more general form
$$V = V_0(r) + r^{-2} V_2 ({\hat {\bf r}})$$
the only restriction on the second term being that it scales under
expansion as $r^{-2}$.  Those familiar with separable systems in 3
dimensions will know that for such potentials $({\textstyle {1
\over 2}} m {\bf h}^2 + V_2)$ is constant along an orbit where for
that case \break $V_2 = V_2 (\theta, \phi)$ and ${\bf h} = {\bf r} \times {\bf
v}$.  The generalisation to 3$N$ dimensions is the first integral
${\textstyle {1 \over 2}} M L^2 + V_2 ({\hat {\bf r}}) = {\textstyle
{1 \over 2}} M {\cal L}^2$ say (note that due to the $V_2$ term,
${\cal L}^2$ does not have to be positive).

The energy equation now reads
$$E = {\textstyle {1 \over 2}} M {\dot {\bf r}}^2 + V_0 + r^{-2} V_2 =
{\textstyle {1 \over 2}} M (\dot r^2 + L^2 r^{-2}) + V_0 + r^{-2} V_2
= {\textstyle {1 \over 2}} M (\dot r ^2 + {\cal L}^2 r^{-2}) + V_0$$
so the $r$ motion pulsates for ever as before.  These systems show no
violent relaxation in their breathing mode which pulsates (or evolves
$E>0$) independently of the complication of the ${\hat {\bf r}}$
motion.  Since $V_2 ({\hat {\bf r}})$ is still free to choose, that
motion can be as complicated as we like to make it.  Defining a new
time $\tau$ by $d/d \tau = 1/r^2 \, d/dt$ the equations of motion for
${\hat {\bf r}}$ as a function of $\tau$ are totally independent of
the $r$ motion, having a reduced Lagrangian system of their own in
$\tau$-time.  An interesting case to consider is the statistical
mechanics of a ``hard cone'' gas in which $V_2$ is large and repulsive
only in very small regions where two particles are nearly in the same
direction as seen from the mass centre.  This corresponds to the small
hard sphere gas so beloved of textbooks.  Carrying out that
statistical mechanics, which is totally independent of any $r$ motion
that may be going on, we obtain a new system at equilibrium in its
${\hat {\bf r}}$ coordinates but pulsating or evolving in $r$.

We have shown (Lynden-Bell \& Lynden-Bell, 1998) this equilibrium to
be best described in terms of the peculiar velocity ${\bf v}_i$
relative to a ``Hubble flow'' $H ({\bf x}_i - {\overline {\bf x}})$
where $H = \dot r/r$ that is
$${\bf v}_i = {\dot {\bf x}}_i - {\dot {\overline {\bf x}}} - H ({\bf
x}_i - {\overline {\bf x}})$$
$$f({\bf v}_i , {\bf x} - {\overline {\bf x}}) \propto \exp \left[ -
\left( {\widetilde {\beta}} r^2 {\textstyle {1 \over 2}} m_i v_i^2 \right)
- {{\widetilde {\beta}} r^2_i \over 2r^2} \right]\ . $$
Thus the distribution is Maxwell-Boltzmann relative to be mean Hubble
flow with a temperature proportional to $r^{-2} (t)$ and the profile
is gaussian with a dispersion proportional to $r(t)$.  It is notable
that the `equilibrium' of the ${\hat {\bf r}}$ coordinates is
maintained throughout the pulsation just as the Planck distribution of
cosmic black-body radiation in the Universe is maintained {\it without
interaction} during the expansion of the Universe.  Thus whether the
relaxation to equilibrium of the angular coordinates is longer than or
shorter than the pulsation time of $r$ is not relevant because
`equilibrium' once attained is maintained throughout the pulsation, it
does not have to be recreated as each radius $r$ is attained.

\end{document}